\documentclass{article}
\usepackage{spconf,amsmath,graphicx}

\usepackage{hyperref}
\usepackage{cite}
\usepackage{amsmath,amssymb,amsfonts}
\usepackage{graphicx}
\usepackage{textcomp}
\usepackage{xcolor}
\def\BibTeX{{\rm B\kern-.05em{\sc i\kern-.025em b}\kern-.08em
    T\kern-.1667em\lower.7ex\hbox{E}\kern-.125emX}}

\usepackage{algorithm}
\usepackage{algpseudocode}
\usepackage{enumitem}

\usepackage{array}
\usepackage{booktabs} 
\usepackage{multirow}
\usepackage{siunitx}

\newcolumntype{L}[1]{>{\raggedright\arraybackslash}p{#1}}
\newcolumntype{C}[1]{>{\centering\arraybackslash}p{#1}}

\newcommand{\C}{\mathbb C}

\newcommand{\R}{\mathbb R}

\newcommand{\Dd}{\mathbf D}
\newcommand{\Rd}{\mathbf{R}}

\newcommand{\sd}{\mathbf{s}}                        

\newcommand{\XX}{\mathbf{x}}

\newcommand{\ZZ}{\mathbf{z}}

\newcommand{\YY}{\mathbf{y}}

\newcommand{\Ad}{\mathbf A}                        
 
\newcommand{\Fd}{\mathbf F} 
\newcommand{\Bd}{\mathbf B}

\newcommand{\Su}{\mathbf{S}_I}

\newcommand{\herm}{{\scriptstyle \boldsymbol{\mathsf{H}}}}
\newcommand{\trans}{{\scriptstyle \boldsymbol{\mathsf{T}}}}

\newcommand{\LLambda}{\boldsymbol{\Lambda}}

\usepackage{multirow}
\usepackage{siunitx}
\usepackage{lipsum}

\usepackage{blindtext}

\title{Learning spatially adaptive sparsity level maps for arbitrary convolutional dictionaries}

\name{Joshua Schulz$^{\star}$, David Schote$^{\star}$, Christoph Kolbitsch$^{\star}$, Kostas Papafitsoros$^{\dagger}$, Andreas Kofler$^{\star}$}
\address{$^\star$Physikalisch-Technische Bundesanstalt (PTB), Braunschweig and Berlin, Germany\\ $^\dagger$School of Mathematical Sciences, Queen Mary University of London, UK}

\begin{document}

\maketitle

\begin{abstract}
State-of-the-art learned reconstruction methods often rely on black-box modules that, despite their strong performance, raise questions about their interpretability and robustness. Here, we build on a recently proposed image reconstruction method, which is based on embedding data-driven information into a model-based convolutional dictionary regularization via neural network-inferred spatially adaptive sparsity level maps. By means of improved network design and dedicated training strategies, we extend the method to achieve filter-permutation invariance as well as the possibility to change the convolutional dictionary at inference time. We apply our method to low-field MRI and compare it to several other recent deep learning-based methods, also on in vivo data, where the benefit of using a different dictionary is demonstrated. We further assess the method’s robustness when tested on in- and out-of-distribution data.  When tested on the latter, the proposed method suffers less from the data distribution shift compared to the other learned methods, which we attribute to its reduced reliance on training data due to its underlying model-based reconstruction component. 
\end{abstract}


\section{Introduction}

Learned reconstruction methods using neural networks nowadays, without a doubt, define the state-of-the-art in image reconstruction \cite{Arridge_Maass_Oektem_Schoenlieb_2019}. Thereby, an important issue is their present black-box character. Often, one must seek a trade-off between empirical/numerical performance and interpretability/transparency, for example, in terms of the convergence guarantees of the derived reconstruction schemes  \cite{mukherjee2023learned}. Additionally, learned reconstruction methods are reported to quite consistently suffer from data-distribution shifts \cite{darestani2021measuring}. To enhance interpretability, it is possible to carefully employ learnable blocks such that the resulting reconstruction can be linked to a model-based variational problem, see e.g.\ \cite{li2020nett}, \cite{kobler2021total}, \cite{pourya2025dealing}, \cite{kofler2025ell1} and references therein.

Dictionary-learning-based methods constitute another family of data-driven reconstruction approaches \cite{garcia2018convolutional} that are highly interpretable, yet so far few studies have connected them to deep learning-based methods to improve their performance \cite{deep_ksvd}.
In \cite{kofler2025ell1},  an image reconstruction method leveraging spatially adaptive sparsity of the image with respect to a pre-trained convolutional dictionary was introduced. The approach consists of an unrolled scheme that can be trained end-to-end from the raw data measurements to the image estimate, where the sparsity level map is estimated by a convolutional neural network (CNN).
The main limitations in \cite{kofler2025ell1} lie in the architectural design of the CNN, which is dictionary-agnostic, meaning that at inference time, it is not possible to use any other dictionary than the one used for training. A change of the dictionary, e.g., \ in terms of the number of filters $K$  - or even only in terms of their order  - results in a loss of reconstruction performance.

Here, we introduce a flexible framework for learning spatially adaptive sparsity level maps that allows for the employment of arbitrary convolutional dictionaries during inference. We do so via a new CNN design customized to this task, training over different numbers of dictionaries and employing efficient dedicated training strategies. More specifically:
\begin{itemize}[noitemsep, leftmargin=0.2cm]
\setlength\itemsep{0.1em}
\item  We condition the CNN on the employed convolutional dictionary and adapt the architecture to achieve dictionary filter permutation invariance and to allow the use of dictionaries with a different number of filters. 
\item The convolutional dictionaries are varied during training, so that the network learns to estimate appropriate sparsity level maps for different dictionaries of arbitrary sizes. 
\item Truncated backpropagation is used during training, which is necessary since the considered reconstruction problem can be very large due to the potential use of a larger number of dictionary filters and larger CNN blocks.
\item We compare against other well-established methods on retrospectively simulated low-field MR data, where we show that our approach is less affected by out-of-distribution shifts. Further, experiments on in vivo data show that our approach, which benefits from using a different dictionary at inference, yields reconstructions that are comparable to the other methods, while at the same time being interpretable. 
\end{itemize}

\section{Methods}

We briefly revise the method in \cite{kofler2025ell1} to present the required concepts. Consider the typical inverse imaging problem 
\begin{equation}\label{eq:forward_problem}
    \YY = \Ad \XX_{\mathrm{true}} + \mathbf{e},
\end{equation}
where $\YY$ denotes the raw measurement data obtained from the (unknown)  image $\XX_{\mathrm{true}}$ through the forward model $\Ad$, and $\mathbf{e}$ denotes additive Gaussian noise. Since we focus on MR image reconstruction, here $\XX_{\mathrm{true}}$ is complex-valued.

\subsection{Reconstruction with Spatially Adaptive Sparsity Level Maps}
\begin{figure*}[t]
    \centering
    \includegraphics[width=0.85\linewidth]{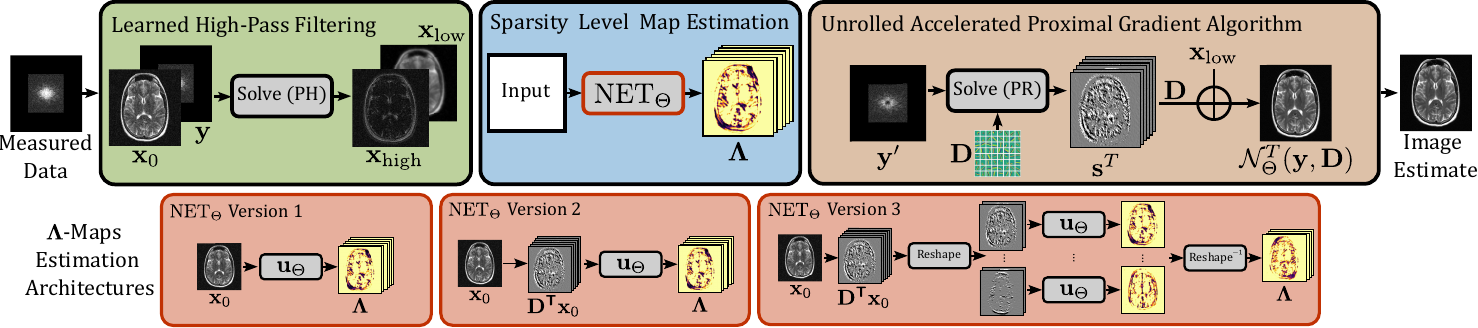}
    \caption{The  reconstruction pipeline of the CDL-$\LLambda$ method and the three different versions of $\mathrm{NET}_{\Theta}$ (\ref{eq:NET_v1}--\ref{eq:NET_v3}) \textit{(orange blocks)} investigated  in this study. First, a learned high-pass filtering step solving \eqref{eq:PL} is performed \textit{(green block)}. Then,  $\mathrm{NET}_{\Theta}$ estimates spatially varying sparsity level maps from an input \textit{(blue block)}. Last, a FISTA algorithm is unrolled to obtain an approximate solution of \eqref{eq:PR} \textit{(ocher block)}. An estimate of the solution is obtained by adding the low-frequency component that was extracted in the first block. 
    In contrast to V1 and V2, the proposed improved V3 is permutation invariant and allows for the use of dictionaries with a different number of filters $K$.}
    \label{fig:pipeline}
\end{figure*}

Reconstructing an estimate of $\XX_{\mathrm{true}}$ by solving \eqref{eq:forward_problem} typically requires the use of regularization methods. 
A successful regularization mechanism can be designed based on spatially adaptive sparsity with respect to a pre-trained convolutional dictionary $\Dd$ \cite{garcia2018convolutional}. Here, one assumes that an image can be represented as a linear combination of sparse feature maps $\{\mathbf{s}_k\}_{k=1}^K$ that are convolved with unit-norm convolutional filters $\{d_k\}_{k=1}^K$.
For conciseness, we use the following notation:
\begin{equation}\label{eq:notation}
    \sd := [\sd_1, \ldots, \sd_K]^\trans,\quad 
    \Dd \sd := \sum_{k=1}^K d_k \ast \sd_k,\quad 
    \Bd := \Ad \Dd.
\end{equation}
As in \cite{wohlberg2016convolutional}, we also aim only to approximate the high-pass-filtered component of the image. Given $\XX_0:=\Ad^\herm \YY$, high-pass filtering of $\XX_0$ is achieved by $\XX_{\mathrm{high}}:= \XX_0 - \XX_{\mathrm{low}}$, where
\begin{equation}
\begin{aligned}
\XX_{\mathrm{low}}&:= 
     \underset{\XX}{\mathrm{argmin}}\;
      \frac{1}{2}\| \XX - \XX_{0}\|_2^2 + \frac{\beta}{2}\| \nabla \XX\|_2^2 \label{eq:PL}
\end{aligned}\tag{PH}
\end{equation}
with $\beta>0$. We strictly enforce the equality $\XX_{\mathrm{high}}=\Dd\sd$ with sparse $\sd$ and  consider the reconstruction problem 
\begin{align}
&\XX^\ast :=\Dd\sd^\ast + \XX_{\mathrm{low}}, \label{eq:res_connection}\\ 
&\sd^\ast:= \underset{\sd}{\arg\min}\,  \frac{1}{2}\|\Bd\sd - \YY^\prime\|_2^2 +  \| \boldsymbol{\LLambda}\, \sd\|_1, \quad \tag{PR}\label{eq:PR}
\end{align}
where $\YY^\prime:=\YY - \Ad\XX_{\mathrm{low}}$ and the sparsity level maps $\LLambda:=[\LLambda_1, \ldots, \LLambda_K]^\trans$ are parametrized as outputs of a deep CNN $\mathrm{NET}_{\Theta}$ with parameters $\Theta$ applied to some input. The method \cite{kofler2025ell1} (named  CDL-$\LLambda$; Convolutional Dictionary Learning with $\LLambda$-maps),  performs the learned high-pass filtering by solving problem \eqref{eq:PL}, the estimation of the sparsity level maps $\LLambda$, and the final reconstruction of the image by solving \eqref{eq:PR} by means of algorithm unrolling \cite{monga2021algorithm} of an accelerated proximal gradient descent (FISTA) \cite{beck2009fast} with convergence guarantees  \cite{chambolle2015convergence}. We denote the entire reconstruction network by $\mathcal{N}_{\Theta}^T$ (see Figure \ref{fig:pipeline}) and the FISTA-scheme to approximately solve \eqref{eq:PR} with $T\in \mathbb{N}$ iterations $\sd^T:= \mathrm{FISTA}^T(\sd_0, \Ad, \Dd, \YY^\prime, \LLambda)$ with starting value $\sd_0=\mathbf{0}\in\C^{NK}$.

\subsection{
Dictionary-Agnostic vs Dictionary-Aware Learning of the Sparsity Level Maps
}
We investigate three different possibilities to construct the network $\mathrm{NET}_{\Theta}$ for estimating spatially adaptive sparsity level maps $\LLambda$, see Figure \ref{fig:pipeline} (bottom) for an overview. Complex-valued images $\XX\in\C^N$ are treated as real-valued two-channel images, and we use real-valued dictionary filters, i.e., 
\[\Dd\sd=\sum_{k=1}^K d_k \ast \sd_k:=\sum_{k=1}^K \big(d_k \ast \mathrm{Re}(\sd_k) + \mathrm{i}\, d_k \ast \mathrm{Im}(\sd_k)\big),
\]
implying that for any $\sd \in \C^{NK}$, and $\ZZ=[\ZZ_1,\ZZ_2]^\trans\in \C^N$
$\Dd \sd= [\Dd, \Dd] [
    \mathrm{Re}(\sd), 
    \mathrm{Im}(\sd)
    ]^\trans,
    \Dd^\trans \ZZ:= [\Dd^\trans \ZZ_1,
    \Dd^\trans \ZZ_2]^\trans.
$
 The weighted $\ell_1$-norm in \eqref{eq:PR} is then defined by $\| \sd\|_1:= \big\| [
    \mathrm{Re} (\sd),  \mathrm{Im} (\sd)]^\trans \big\|_1.$ In \cite{kofler2025ell1}, $\mathrm{NET}_{\Theta}$ was constructed as a U-Net that estimates $K$ sparsity level maps $\LLambda:=[\LLambda_1, \ldots, \LLambda_K]$ (shared between real and imaginary parts of $\sd$) from the input image, i.e., upon vectorization of 2D images to vectors, it holds $\LLambda = \mathrm{NET}_{\Theta}(\XX_0)$ with
\begin{equation}\label{eq:NET_v1}
    \LLambda_k:=\big(\mathrm{NET}_{\Theta}(\XX_0)\big)_k:= t\cdot\boldsymbol{\phi}\, \circ \left(\mathbf{u}_{\Theta} \tag{V1}
    (\XX_0)\right)_k,
\end{equation}
with global learnable scalar parameter $t>0$, Softplus activation function $\boldsymbol{\phi}$ and a 2-to-$K$ 2D U-Net $\mathbf{u}_{\Theta}$.
While proven effective, it is evident from \eqref{eq:NET_v1} that the network is agnostic with respect to the dictionary $\mathbf{D}$ as well as only applicable to dictionaries with a fixed number of filters $K$. We refer to the CNN architecture in \eqref{eq:NET_v1} as " $\mathrm{NET}_{\mathrm{\Theta}}$ V1".

A straightforward solution to condition the network $\mathrm{NET}_{\Theta}$ on the dictionary $\Dd$ is to provide a dictionary-dependent input and use a $K$-to-$K$ 2D U-Net $\mathbf{u}_{\Theta}$  to obtain  
\begin{equation}\label{eq:NET_v2}
    \LLambda:=\mathrm{NET}_{\Theta}(\Dd^\trans\XX_0):= t\cdot\boldsymbol{\phi}\, \circ \mathbf{u}_{\Theta} \tag{V2}
    (\Dd^\trans\XX_0),
\end{equation}
which we refer to as "$\mathrm{NET}_{\mathrm{\Theta}}$ V2". Although using a dictionary-dependent input, $\mathrm{NET}_{\mathrm{\Theta}}$ V2 is nevertheless inherently linked to the number of filters $K$. 

Thus, we consider the third variant "$\mathrm{NET}_{\mathrm{\Theta}}$ V3", which, by slight abuse of notation, reads as
\begin{equation}\label{eq:NET_v3}
    \LLambda:=\mathrm{NET}_{\Theta}(\Dd^\trans\XX_0):= \Rd^{-1} t\cdot\boldsymbol{\phi}\, \circ \mathbf{u}_{\Theta} \tag{V3}
    ( \Rd \Dd^\trans\XX_0),
\end{equation}
where $\Rd$ and $\Rd^{-1}$ are operators that reshape the input tensors by moving the channel dimension to the batch-dimension, i.e., they transform inputs with shapes $(\mathrm{batch}, 2\cdot K, N_y, N_x)$ to $(\mathrm{batch} \cdot K, 2, N_y, N_x)$ and from $(\mathrm{batch} \cdot K, 1, N_y, N_x)$ to $(\mathrm{batch}, K, N_y, N_x)$, respectively. Here, $\mathbf{u}_{\Theta}$ denotes a 2-to-1 2D U-Net. In contrast to V1 and V2, the \textit{same} 2D U-Net is employed to estimate the corresponding sparsity level map $\LLambda_k\in\R_{>0}^N$ from the input $(\Dd^\trans \XX_0)_k \in \R^{2N}$. By doing so, $\mathrm{NET}_{\Theta}$ can be employed with dictionaries with an arbitrary number of filters $K$ at inference.

\subsection{Network Training}
Instead of a single, fixed dictionary, 
here, we employ an entire set of convolutional dictionaries $\Dd$ with different numbers of filters $K$ and kernel sizes $k_f \times k_f$. Given training data, we minimize a mean squared error (MSE)-based loss function 
\begin{equation}\label{eq:loss_function}
    \mathcal{L}(\Theta):= \sum_{\YY,\, \XX_{\mathrm{true}},\, \mathbf{D}} \mathrm{MSE} \big(\mathcal{N}_{\Theta}^T(\YY, \Dd)\,,\, \XX_{\mathrm{true}}\big),
\end{equation}
which means that the network $\mathrm{NET}_{\Theta}$ is exposed not only to various raw data-target pairs, but also to different reconstruction problems of the form \eqref{eq:PR}.

Being able to employ larger dictionaries in terms of larger $K$ and kernel size $k_f \times k_f$ with richer structure implies increased memory consumption when unrolling the FISTA-block. 
Therefore, employing truncated back-propagation \cite{shaban2019truncated} is an attractive choice for unrolling a relatively large number of iterations with many large intermediate quantities. More precisely, during training,  we first compute $\sd^{T^\prime} = \mathrm{FISTA}^{T^\prime}(\sd_0, \Ad,\Dd, \YY^{\prime}, \LLambda)$ with the FISTA-block with the current estimate of $\LLambda$ \textit{without} tracking gradients. Then, we perform additional $T-T^{\prime}$ unrolled iterations (starting the FISTA iteration from $\sd^{T^\prime}$) by also tracking the gradients of $\Theta$ to train by minimizing \eqref{eq:loss_function}. We used $T=64$ and $T^{\prime}=36$.

\subsection{Datasets and Evaluation Metrics}\label{subsec:dataset_n_eval}
Here, we conduct experiments on a low-field (LF) MR reconstruction problem. 
LF MRI suffers from high noise and low image resolution due to hardware constraints. We therefore model the forward operator $\Ad$ in \eqref{eq:forward_problem} by $\Ad:=\Su \Fd$, where $\Fd$ denotes a 2D Fourier transform and $\Su$ a binary mask that masks out the outer region of the measurements $\YY$, modeling the low-resolution acquisition. 
We used brain MR (4875/1393/696 images for training/validation/testing) and 3951 knee MR images from the fastMRI dataset \cite{zbontar2018fastmri} to retrospectively simulate data according to \eqref{eq:forward_problem}. The variance of the noise component $\mathbf{e}$ in \eqref{eq:forward_problem} is set relatively high by $\sigma^2 \in \{0.2, 0.3\}$. Further, we show an application of the methods to an in vivo T2-weighted brain image acquired with the Open Source Imaging Initiative (OSI$^2$) LF MR scanner \cite{OSI2}.
We report two distortion-based metrics, i.e., the MSE and the structural similarity index measure (SSIM), as well as the reference-free blur image metric \cite{crete2007blur}, which is especially informative for the in vivo data, where no target image exists.
All metrics are computed by discarding the background of the images, employing masks derived from the target images. The MSE and SSIM were calculated using the \texttt{MRpro} library \cite{zimmermann2025}, which allows for the definition of such masks, while the blur metric implementation is from \texttt{Scikit-learn} \cite{kramer2016scikit}. 
For MoDL, E2E VarNet, SRDenseNet, and CDL-$\LLambda$, we used a learning rate of $10^{-4}$ for the employed CNN-blocks, and $10^{-2}$ for the learned step sizes/scalars. CDL-$\LLambda$ was trained for 48 epochs using a batch size of 1, the others for 128 epochs using a batch size of 4. Code and further training details are available at 
\url{https://github.com/bjpschulz/AdaConvSynth}.

\subsection{Pre-Training the Dictionaries}
We pre-trained different convolutional dictionaries with different numbers of filters $K$ of kernel sizes $k_f\times k_f$. For doing that, we employed the sparse coding library SPORCO \cite{wohlberg2017sporco} to solve 
a convolutional dictionary learning problem using 360 brain MR images using the online dictionary learning method described in \cite{liu2018first} for different choices of $\beta=0.1, 0.25, 0.5$ in \eqref{eq:PL},  scalar sparsity level parameters $\lambda = 0.1, 0.5, 2, 4$, kernel sizes $k_f=9,11$ and number of filters $K=16,32,64,128$, resulting in an overall number of 96 different dictionaries.
Note that we do not use the $K=128$-dictionaries when minimizing \eqref{eq:loss_function} due to GPU-memory constraints.

\subsection{Methods of Comparison}
We compare against the End-to-End Variational Network (E2E VarNet) \cite{sriram2020end}, the model-based deep learning (MoDL) \cite{aggarwal2018modl}, and a low-field MRI super-resolution (SR) approach, SRDenseNet \cite{de2022deep}. The first two methods both employ the measured data $\YY$, while the latter maps the IFFT reconstruction obtained from the $160\times160$ $k$-space data to a $320 \times 320$-sized image. 
Our implementations of MoDL, E2E VarNet and SRDenseNet contain $113\,413$, $34\,512\,846$ and $5\,654\,834$ trainable parameters. MoDL is unrolled for $T=10$ iterations and alternates between the application of the CNN-block and a data-consistency (DC) step involving a learnable scalar parameter $\lambda>0$  that is shared among the iterations. E2E VarNet is unrolled with $T=12$ iterations and contains iteration-dependent learned step-sizes.
MoDL and E2E VarNet were accordingly adapted to the here considered single-coil case, i.e., the DC step in MoDL has a closed-form solution instead of requiring a conjugate gradient method, and E2E VarNet does not employ a coil-sensitivity map-refinement module.
CDL-$\LLambda$ contains $8\,266\,914$ and $2\,104\,802$ trainable parameters for V1 and V2, respectively, while for  V3, we use a more shallow U-Net with only $112\,067$ trainable parameters (dictionary filters not included).

\section{Results}
\begin{figure}[h]
    \centering
    \includegraphics[width=\linewidth]{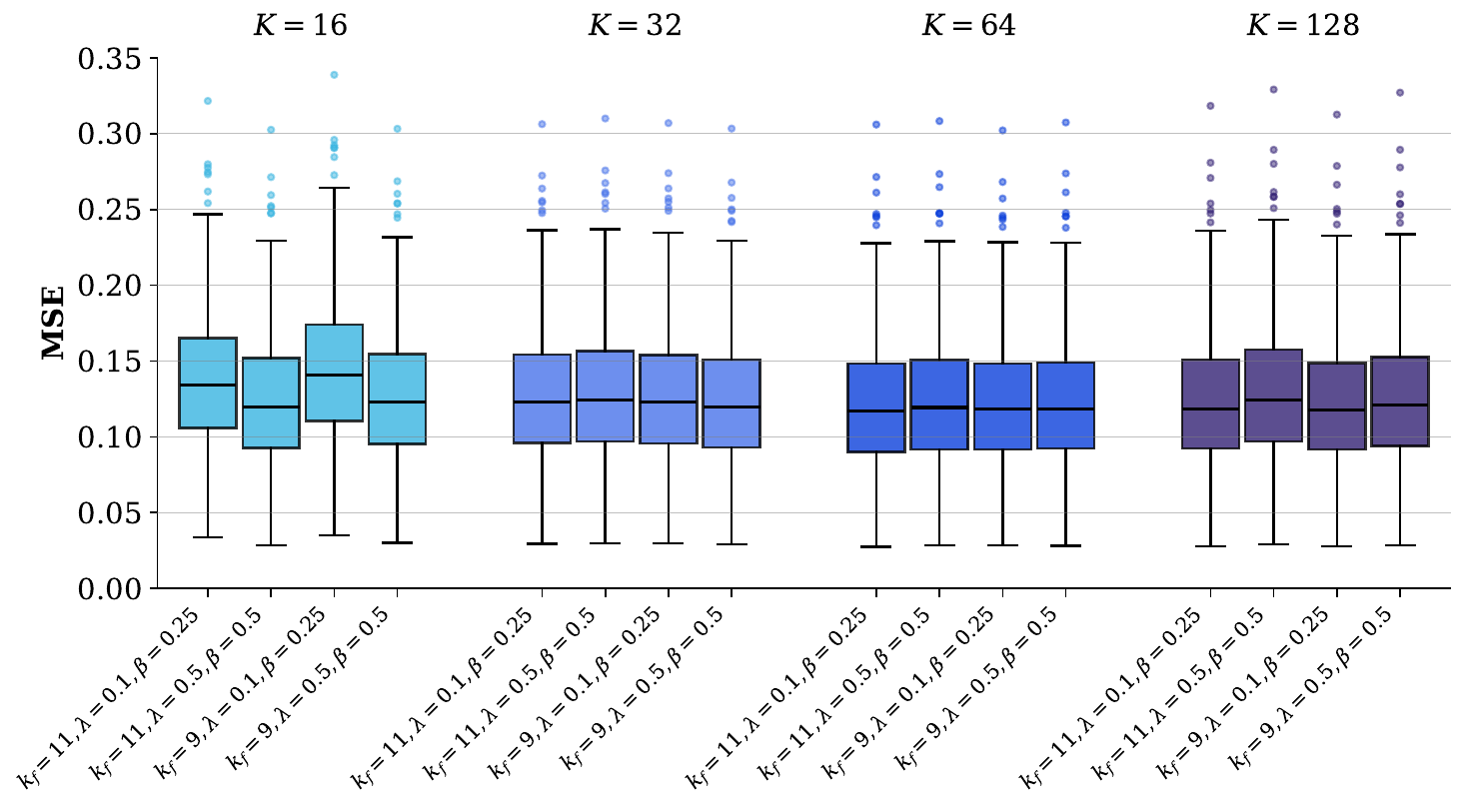}
    \caption{MSE obtained by CDL-$\LLambda$ with $\mathrm{NET}_{\Theta}$ V3 over the brain MR test set for 16 different choices of dictionaries. Note that, for training, we did not use the $K=128$-dictionaries.}
    \label{fig:changing_dictionary_at_inference}
\end{figure}

\textit{Dictionary Filter-Permutation Invariance and Different Numbers of Filters $K$:}
Table  \ref{tab:permutation_results} shows the change in SSIM and MSE when the three different versions of $\mathrm{NET}_{\Theta}$ described in \eqref{eq:NET_v1}, \eqref{eq:NET_v2}, and \eqref{eq:NET_v3} are exposed to a permutation of the dictionary filters. In this case, training was carried out using only one dictionary $\mathbf{D}$ with $K=32$ filters of size $k_f \times k_f = 11 \times 11$. All three versions achieve approximately the same performance in terms of SSIM and MSE, while V3 is the only one invariant under filter order permutation. Additionally,  when training the improved $\mathrm{NET}_{\Theta}$ V3 according to \eqref{eq:loss_function}, it is possible to employ a different dictionary $\mathbf{D}$ at inference by maintaining a comparable performance, see Figure \ref{fig:changing_dictionary_at_inference}.

\begin{table}[t]
\centering
\small
\renewcommand{\arraystretch}{1.1}
\setlength{\tabcolsep}{6pt}
\begin{tabular}{
  L{6mm}  
  |L{10mm} 
  |S[table-format=1.2]@{$\pm$}S[table-format=1.2]
  |S[table-format=1.2]@{$\pm$}S[table-format=1.2]
  |S[table-format=1.2]@{$\pm$}S[table-format=1.2]
}
\hline
& & \multicolumn{2}{c|}{\textbf{V1}} & \multicolumn{2}{c|}{\textbf{V2}} & \multicolumn{2}{c}{\textbf{V3}} \\
\multirow{2}{*}{Base}
& SSIM & 0.85 &  0.05 & 0.82  & 0.05 & 0.82  & 0.05 \\
& MSE  & 0.11 &  0.04 & 0.14  & 0.06 & 0.14 & 0.06 \\
\hline
\multirow{2}{*}{$\pi_1$}
& $\Delta$SSIM & -0.06 &  0.01 & -0.08  & 0.03 &  \multicolumn{2}{c}{0}  \\
& $\Delta$MSE  & +0.12 &  0.04 & +0.14  & 0.10 &  \multicolumn{2}{c}{0}  \\
\hline
\multirow{2}{*}{$\pi_2$}
& $\Delta$SSIM & -0.05 &  0.01 & -0.06  & 0.02  & \multicolumn{2}{c}{0}  \\
& $\Delta$MSE  & +0.12 &  0.04 & +0.12  & 0.08 & \multicolumn{2}{c}{0}  \\
\hline
\multirow{2}{*}{$\pi_3$}
& $\Delta$SSIM & -0.05 &  0.01 & -0.07  & 0.02 & \multicolumn{2}{c}{0}   \\
& $\Delta$MSE  & +0.11 &  0.04 & +0.17  & 0.11 & \multicolumn{2}{c}{0}   \\
\hline
\end{tabular}
\caption{Results for $\mathrm{NET}_{\Theta}$ as in \eqref{eq:NET_v1}, \eqref{eq:NET_v2} and \eqref{eq:NET_v3} for three filter permutations $\pi_1,\pi_2,\pi_3$, shown for $K=32, 11\times 11$-kernels. The improved V3 is filter permutation invariant by construction. Note that differences between the baselines of V1, V2, and V3 can be attributed to the deviations of the three versions in terms of the number of trainable parameters, which can be easily addressed by increasing them.}
\label{tab:permutation_results}
\end{table}

\begin{figure}
    \centering
    \includegraphics[width=\linewidth]{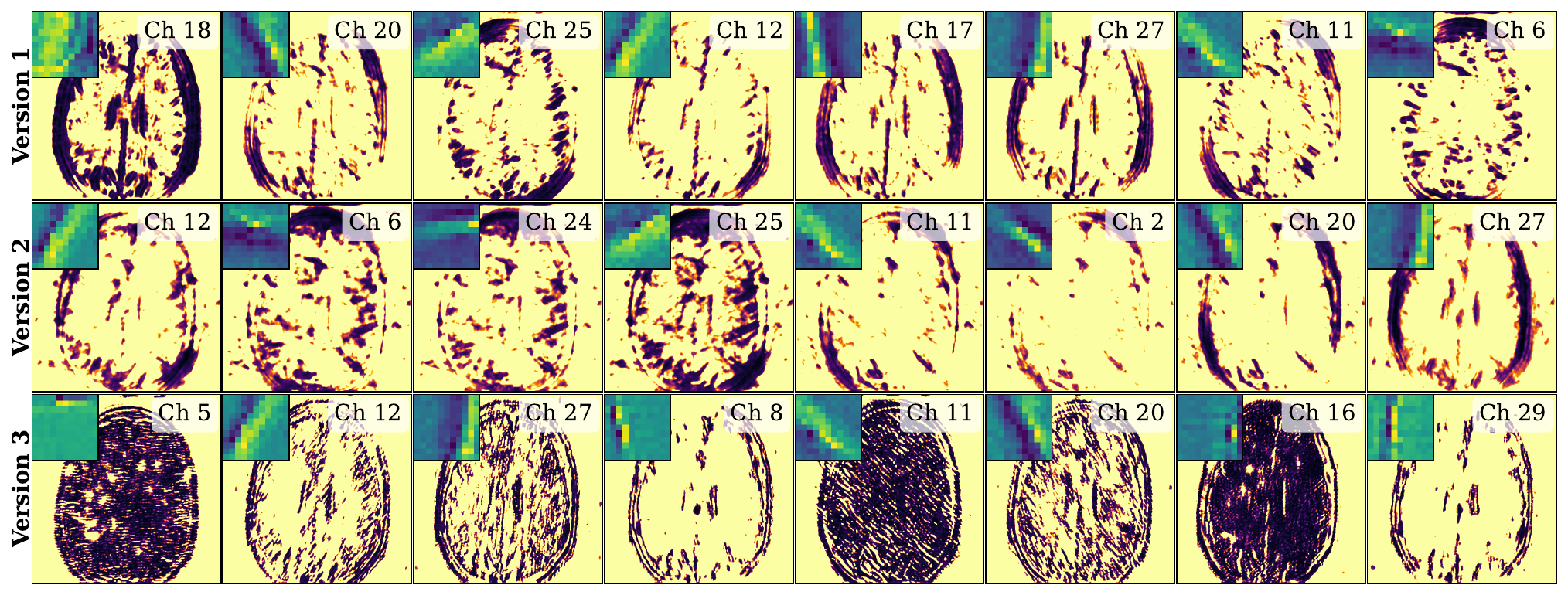}
    \caption{Eight out of $K=32$  $\LLambda$-maps with the largest variance,  seen as indicative of the filter's importance in the image representation,  for each $\mathrm{NET}_{\Theta}$ V1, V2, V3, and  for filter size $k_f \times k_f = 11 \times 11$.}
    \label{fig:lambda_maps_comparison}
\end{figure}

\textit{Inspection of the Sparsity Level Maps:}
Figure \ref{fig:lambda_maps_comparison} shows an exemplary comparison of eight  $\LLambda$-maps for $\mathrm{NET}_{\Theta}$ \eqref{eq:NET_v1}-\eqref{eq:NET_v3} for a dictionary with $K=32$ filters of size $k_f\times k_f=11 \times 11$. The shown maps are those with the largest variances relative to the entire set of $\LLambda$-maps for each version. The figure confirms that all three versions of $\mathrm{NET}_{\Theta}$ assign similar importance (indicated by the larger variance of the corresponding map, see also \cite{kofler2025ell1}) to the respective filter. For example, filter 25 in V1 appears to contribute to the representation of the image also in V2, filter 12 and 20 in all three versions, etc.

\begin{figure}[t]
    \centering
    \begin{minipage}[t]{0.5\linewidth}
        \centering
        \includegraphics[height=9.8cm,keepaspectratio]{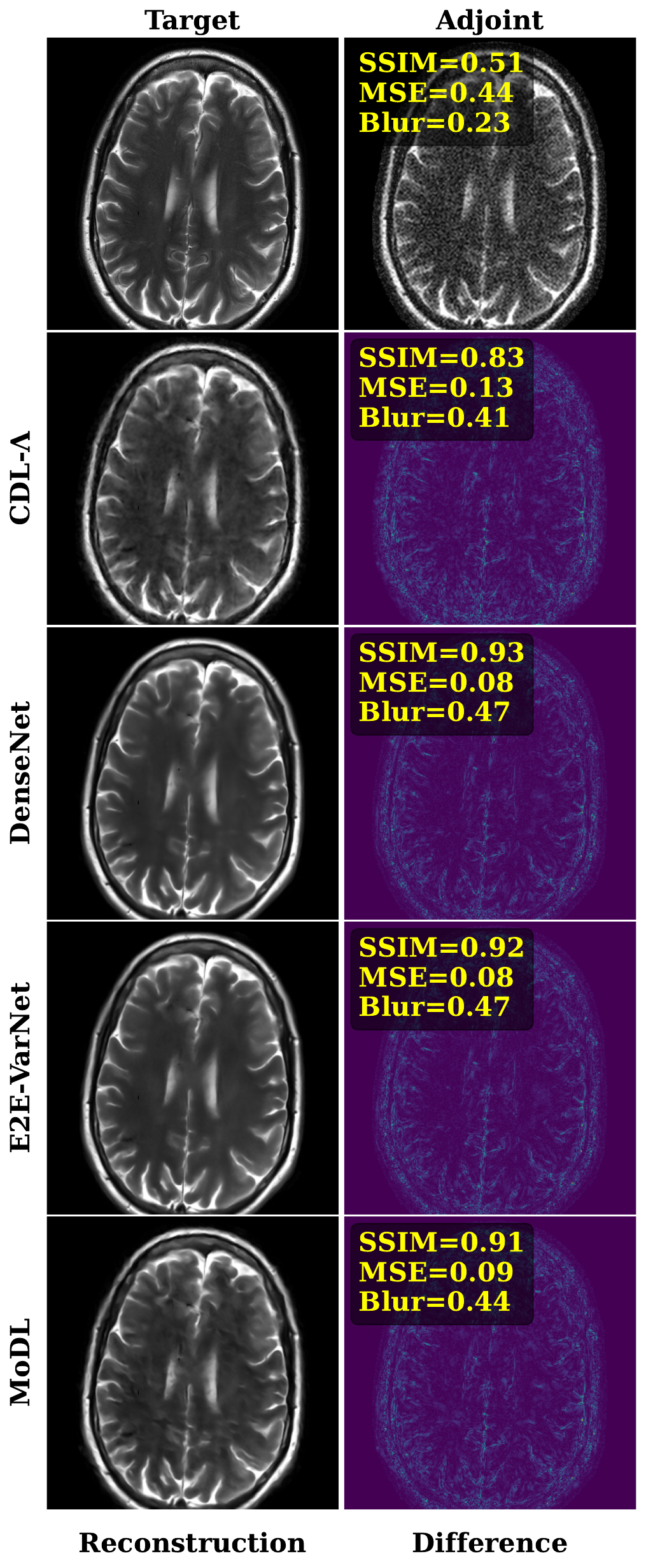}
    \end{minipage}%
    \hfill%
    \begin{minipage}[t]{0.5\linewidth}
        \centering
    \includegraphics[height=9.8cm,keepaspectratio]{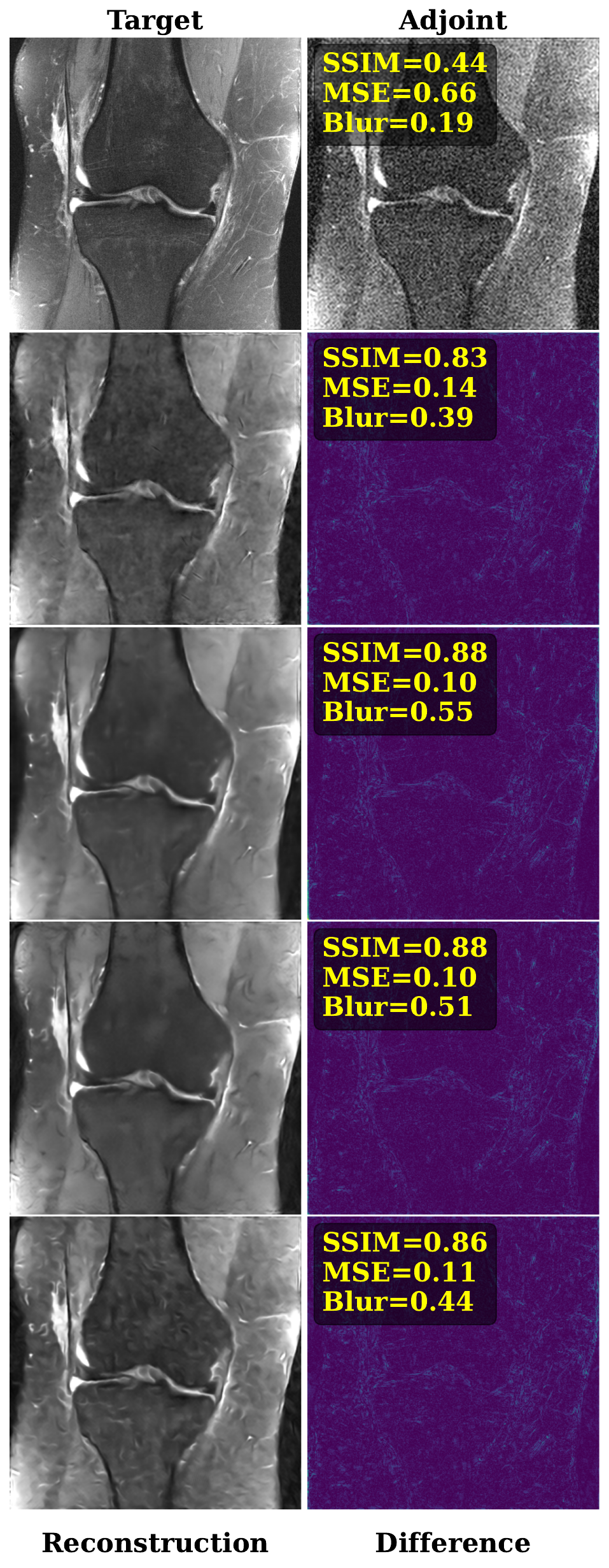}
    \end{minipage}
    \caption{A comparison of MoDL \cite{aggarwal2018modl}, E2E VarNet \cite{sriram2020end}, SRDenseNet \cite{de2022deep} and CDL-$\LLambda$ ($K=64, 11\times 11$-kernels) on in-distribution brain MR data (left, $\sigma^2=0.3$) and out-of-distribution knee MR data (right, $\sigma^2=0.2$).}
    \label{fig:recon_results}
\end{figure}

\textit{Comparison to Other Methods:}
Figure \ref{fig:recon_results} shows a visual comparison of the adjoint reconstruction, MoDL, E2E VarNet, SRDenseNet, and the investigated CDL-$\LLambda$ using $K=64$ filters of size $k_f \times k_f = 11 \times 11$ for a brain and knee MR image, together with their respective error images and image metrics.
\begin{figure}[h!]
    \centering
    \includegraphics[width=0.99\linewidth]{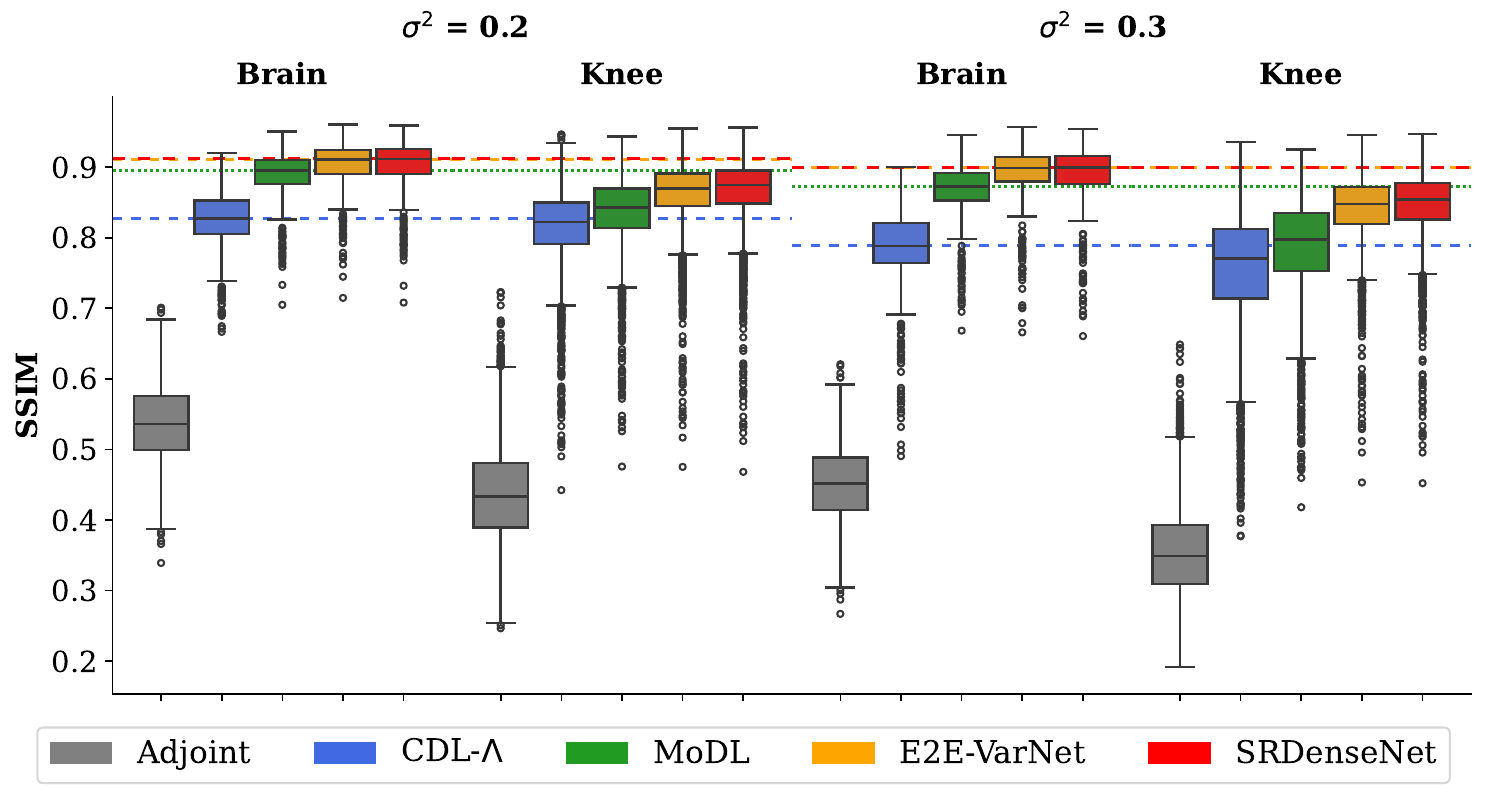}
    \caption{SSIM scores on the brain and knee MR test datasets for noise variances $\sigma^2=0.2$ and $\sigma^2=0.3$. The dashed lines denote the median SSIM value of the respective methods on the brain dataset. Note how for MoDL, E2E-VarNet, and SRDenseNet, the median brain-SSIM lies outside the interquartile range of the corresponding knee-SSIM distributions.
    }
    \label{fig:ood_measures}
\end{figure}
\begin{figure}[h!]
    \centering
    \includegraphics[width=0.96\linewidth]{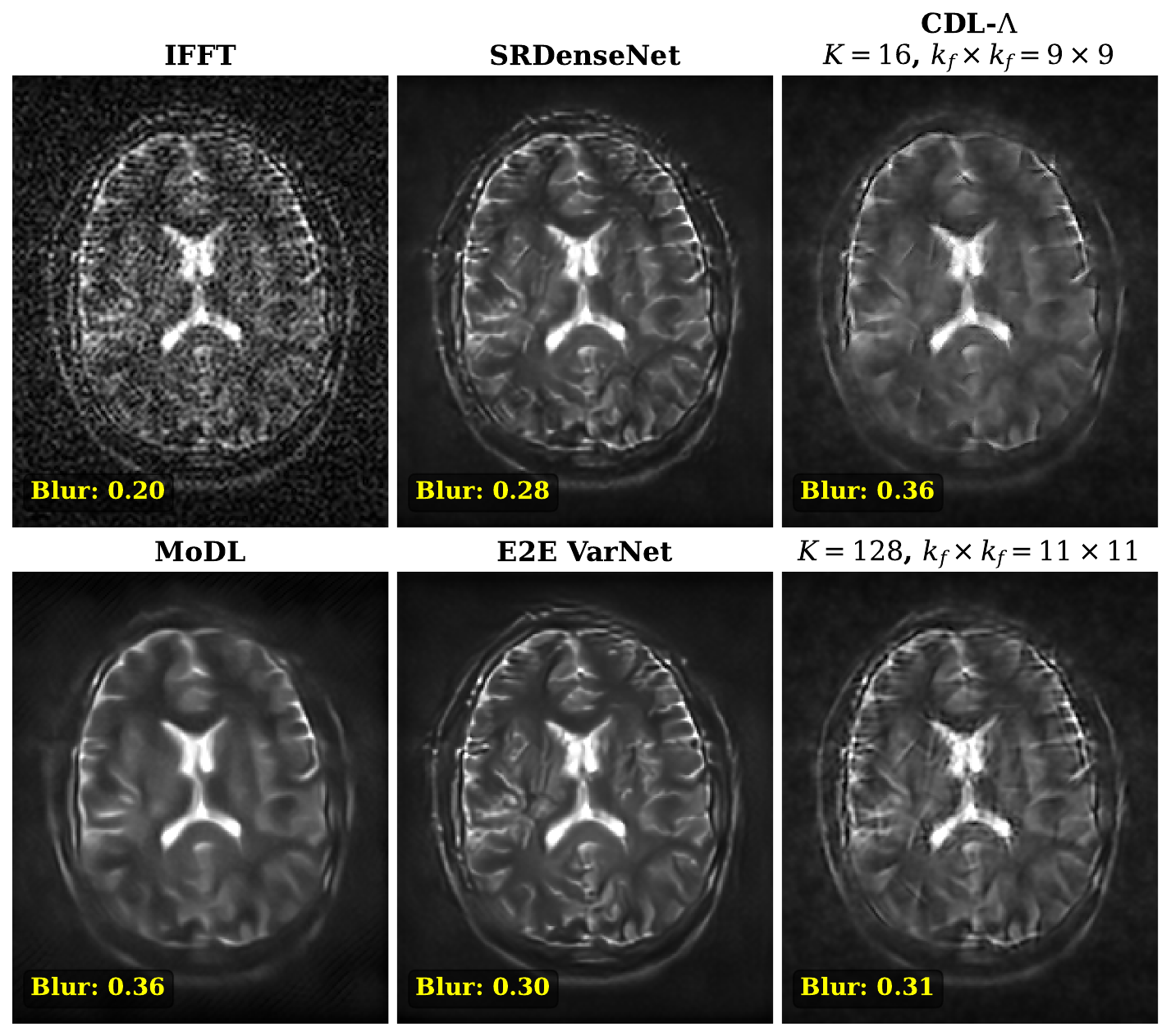}
    \caption{In vivo T2-weighted brain MR image: MoDL, E2E VarNet  and SRDenseNet, and the  CDL-$\LLambda$  with the improved  $\mathrm{NET}_{\Theta}$ \eqref{eq:NET_v3}, once with $K=16$ filters of size $k_f\times k_f = 9 \times 9$, and once with $K=128$, $k_f\times k_f = 11 \times 11$. Note how the use of a larger dictionary at inference, which was not used in training, leads to a sharper result.
    } 
    \label{fig:results_osi2}
\end{figure} 
Despite yielding accurate reconstructions, CDL-$\LLambda$ is surpassed by all other learned methods. However, when tested on out-of-distribution data, i.e., on the knee images, the performance gap between the different methods is noticeably reduced.
For example, compare the reduced difference between MoDL and CDL-$\LLambda$ in terms of SSIM for brain vs knee images, for $\sigma^2=0.2$ and $\sigma^2=0.3$ in Figure \ref{fig:ood_measures}.

\textit{Application to In Vivo Data:}
Figure \ref{fig:results_osi2} shows a comparison of the methods applied to a T2-weighted in vivo image. Since no target image is available, the blur metric (the higher, the blurrier) is displayed together with the reconstructed image. All methods successfully removed noise and improved the resolution, producing comparable results, with MoDL exhibiting a slight tendency to oversmooth image details. Observe also how CDL-$\LLambda$ in this case benefits from using a larger dictionary at inference ($K=128$) that leads to a sharper result.

\section{Conclusion}
Together with the improved CNN-block presented here, CDL-$\LLambda$  yields an interpretable sparsity-based method with convergence guarantees that is less prone to suffering from data distribution shifts compared to MoDL, E2E VarNet, and SRDenseNet. This can be explained by the reduced reliance on training data due to its underlying model-based reconstruction component. Indeed, the CNN-block $\mathrm{NET}_{\Theta}$ in CDL-$\LLambda$,  solely estimates the $\LLambda$-maps, and the regularizing effect comes from the sparsity imposed in \eqref{eq:PR}, whereas in the   other approaches, deep CNNs are directly responsible for the reduction of noise and artefacts.

Furthermore, with the improved CNN-block $\mathrm{NET}_{\Theta}$ V3, CDL-$\LLambda$ can now be equipped with a general sparsity level map estimator for arbitrary convolutional dictionaries. We showed that thanks to the use of a larger dictionary at inference, competitive performance can be achieved for in vivo data as well. Our work paves the way to be able to utilize the estimated sparsity level maps to further adapt the dictionary filters as well as to develop rejection and/or replacement strategies for less useful dictionary filters for the image representation, possibly in a zero-shot self-supervised fashion.

\vfill\pagebreak

\bibliographystyle{IEEEbib}
\bibliography{refs}

\end{document}